\documentclass[12pt,twoside]{article}
\usepackage[mathscr]{eucal}
\usepackage{amsmath,amsfonts,amssymb,amsthm,mathabx,empheq}
\usepackage{times}
\usepackage{pdfsync}
\usepackage{cite}
\usepackage{url}
\usepackage{hyperref}
\usepackage{tensor}
\usepackage{color}
\usepackage{multicol}
\usepackage{bbold}

\advance\textheight\topskip
\allowdisplaybreaks[1]

\newcommand\td{\text{d}}
\newcommand\cO{{\cal O}}

\newcommand{\p}{\partial}

\newcommand{\be}{\begin{equation}}
\newcommand{\ee}{\end{equation}}
\newcommand{\bea}{\begin{eqnarray}}
\newcommand{\eea}{\end{eqnarray}}

\def\bz{\bar z}

\def\n{\nabla}
\def\bm{\bar{m}}

\newcommand{\nn}{\nonumber}
\newcommand*\xbar[1]{%
  \hbox{%
    \vbox{%
      \hrule height 0.5pt 
      \kern0.3ex
      \hbox{%
        \kern-0.0em
        \ensuremath{#1}%
        \kern-0.0em
      }%
    }%
  }%
}

\allowdisplaybreaks[1]

\DeclareFontFamily{OT1}{rsfs}{} \DeclareFontShape{OT1}{rsfs}{m}{n}{
<-7> rsfs5 <7-10> rsfs7 <10-> rsfs10}{}
\DeclareMathAlphabet{\mycal}{OT1}{rsfs}{m}{n}

\hypersetup{
colorlinks=true,
linktoc=page,    
linkcolor=blue,
citecolor=blue,
urlcolor=blue,
colorlinks=true,
}

\begin{document}

\title{Note on identifying four-dimensional vacuum solutions from Weyl invariants}

\author{Pujian Mao}
\date{}

\def\mytitle{Note on identifying four-dimensional vacuum solutions from Weyl invariants}

\addtolength{\headsep}{4pt}

\begin{centering}

  \vspace{1cm}

  \textbf{\Large{\mytitle}}

  \vspace{1cm}

  {\large Pujian Mao }

\vspace{.5cm}

\vspace{.5cm}
\begin{minipage}{.9\textwidth}\small \it  \begin{center}
     Center for Joint Quantum Studies, Department of Physics,\\
     School of Science, Tianjin University, 135 Yaguan Road, Tianjin 300350, China
 \end{center}
\end{minipage}

\end{centering}

\begin{center}
Emails: pjmao@tju.edu.cn
\end{center}

\begin{center}
\begin{minipage}{.9\textwidth}
\textsc{Abstract}: The diffeomorphism covariance is a fundamental property of General Relativity which leads to the fact that the same solution of Einstein equation can be given in completely distinct forms in different coordinate systems. Distinguishing or identifying two metrics as solutions of Einstein equation is particularly challenging. In a recent paper \cite{Lu:2025fzm}, it is proposed to apply the relations of different Weyl invariants to distinguish solutions. In this note, we present a complementary application of the Weyl invariants. We verify from Weyl invariants that two metrics with completely different forms are the same solution. We also present the coordinates transformation that connects the two metrics.

\end{minipage}
\end{center}

\thispagestyle{empty}



\section{Introduction}

General Relativity (GR) is a fundamental theory of modern physics. One of the profoundest significances of GR is that it satisfies the general principle of relativity, which states that the laws of physics are the same for all observers and there will never be a preferred reference. The core of GR is the Einstein equation, which describes the relation between the geometry of spacetime and the matter fields that are contained in the spacetime. From the point of view of the Einstein equation, the general principle of relativity is resided in the diffeomorphism covariance. The Einstein equation is constructed from a metric. However, the same solution of Einstein equation can be given in completely distinct metric forms in different coordinates because of the diffeomorphism covariance. Thus distinguishing solutions of Einstein equation is of intrinsic significance in GR. 

The natural proposal to distinguish solutions is to check coordinate-invariant properties. For instance, the algebraic classification of the Weyl tensor \cite{Petrov:2000bs} provides a very powerful way to classify exact solutions. In the same type of Petrov class, one can check the relations of Weyl invariants, in particular for the vacuum solutions. In four dimensions, a simple example to distinguish Schwarzschild solution from most of other type-D solutions is that the square of the Weyl tensor $W_2=W^{\mu\nu\alpha\beta}W_{\mu\nu\alpha\beta}$ and the cubic of the Weyl tensor $W_3=W^{\mu\nu\alpha\beta}W_{\mu\nu\rho\sigma}{W_{\alpha\beta}}^{\rho\sigma}$ satisfy a unique relation, namely $(W_2)^3=12(W_3)^2$. Thus, any solution that does not satisfy the relation $(W_2)^3=12(W_3)^2$ must not be Schwarzschild. Though the relations of Weyl invariants can decisively distinguish different solutions, the computation of various Weyl invariants can be very complicated in general. Nevertheless, it is shown that \cite{Lu:2025fzm} the derivation of Weyl invariants can be performed very efficiently in the Newman-Penrose (NP) formalism \cite{Newman:1961qr}, see also earlier investigations, e.g., in \cite{Cherubini:2002gen}. Subsequently, a large class of Weyl invariants are calculated in \cite{Lu:2025fzm} to distinguish an infinite tower of algebraically special vacuum solutions. 

Weyl invariants provide a generally applicable tool for distinguishing solutions in GR. However, there is still the converse problem that how can one identify two distinct metrics if they represent the same physical solution. The most direct approach is to find the explicit coordinate transformation relating the two metrics, but this is often highly nontrivial. In fact, deriving such a transformation can be even more complicated than solving Einstein equation itself.

In this note, we present an example illustrating how Weyl invariants can be used to identify two metrics as the same vacuum solution of Einstein equation. Since any difference from the relation of Weyl invariants uniquely distinguishes two distinct solutions, verifying equivalence through Weyl invariants requires, in principle, the comparison of all possible Weyl invariants. This task is in general very challenging. Fortunately, in certain special cases, one can demonstrate that all possible relations among the Weyl invariants of two metrics coincide, thereby proving that the two metrics describe the same solution in different coordinate systems. The verification in this work is revealed in the NP formalism. For notation and conventions of the NP formalism, we would refer to \cite{Chandrasekhar}, and for additional information see also \cite{Lu:2025fzm}.

\section{New metric from Hopf structure}

A new metric is constructed from the Kerr-Schild form in \cite{Harada:2025krm} from the Hopf structure \cite{Hopf1931}. The Kerr-Schild form of a spacetime metric is the assumption that the full metric can be written as a direct sum of a flat spacetime metric and a product of a scalar field and two null vectors. More precisely, the line-element takes the form
\be
\td s^2=\eta_{\mu\nu} \td x^\mu \td x^\nu + H k_\mu k_\nu \td x^\mu \td x^\nu,
\ee
where $H$ is a real scalar function and the vector $k_\mu$ is null with respect to both the full spacetime metric and the flat spacetime metric $\eta_{\mu\nu}$. The remarkable discovery in \cite{Harada:2025krm} is that the null vector $k_\mu$ can be constructed from a Hopf fibration \cite{Hopf1931}. Consider two complex numbers $z_1, z_2 \in \mathbb{C}$ satisfying $|z_1|^2 + |z_2|^2 = 1$. They can be parametrized as
\begin{equation}
z_1 = \frac{2(x_2 + i x_3)}{r^2 - t^2 + 1 + 2 i t}, \qquad
z_2 = \frac{r^2 - t^2 - 1 + 2 i t}{r^2 - t^2 + 1 + 2 i t},
\label{eq:Hopf_z}
\end{equation}
where $r^2 = x_1^2 + x_2^2 + x_3^2$. Such pair of complex numbers $(z_1, z_2)$ defines coordinates on the 3-sphere $S^3$. At each point on $S^3$, the Hopf mapping
\begin{equation}
(z_1, z_2) \rightarrow \frac{z_1}{z_2} = \frac{2(t + i x_2)}{r^2 - t^2 - 1 - 2 i x_1} \equiv f(t, x_1,x_2,x_3),
\label{eq:Hopf_map}
\end{equation}
can be interpreted as a map from 3-shpere to a 2-sphere, where $f$ defines a complex scalar. Since $z_1 / z_2 = e^{i \alpha} z_1/ (e^{i \alpha} z_2)$ for $\alpha = [0, 2\pi)$, all points on the circle $e^{i\alpha} (z_1,z_2)$ in $S^3$ map to the same point $f = z_1 / z_2$ on $S^2$. Therefore, the inverse map from $z_1 / z_2$ on $S^2$ to the points on $S^3$ forms an $S^1$ fiber \cite{Hopf1931}. The salient feature of the gravitational Hopf structure is that the vector $k_\mu$ in the Kerr-Schild form is tangent to a Hopf fibration on all constant-time spatial foliations of Minkowski space. In a covariant description, a tangent vector $k^\mu$ to this $S^1$ fiber satisfies \cite{Harada:2025krm}
\begin{equation}
A_{\mu\nu} k^\nu = 0,\qquad
A_{\mu\nu} = \frac{1}{2i} (\partial_\mu f^* \, \partial_\nu f - \partial_\nu f^* \, \partial_\mu f).
\label{eq:Fmunu}
\end{equation}
Note that the vector $k_\mu$ in the Kerr-Schild form also satisfies both the null condition $\eta^{\mu\nu} k_\mu k_\nu = 0$ and the geodesic condition $k^\nu \partial_\nu k_\mu = 0$. A solution of $k_\mu$ with respect to these conditions is obtained in \cite{Harada:2025krm}, which is given by
\begin{equation}
k_\mu = \frac{1}{\sqrt{2}(1 + (t - x_1)^2)} 
\begin{pmatrix}
1 + (t - x_1)^2 + x_2^2 + x_3^2 \\
1 + (t - x_1)^2 - x_2^2 - x_3^2\\
-2(t - x_1) x_2 - 2x_3 \\
-2(t - x_1) x_3 + 2x_2 
\end{pmatrix}.
\end{equation}
Since $k_\mu$ depends on $t - x_1$, it is convenient to adopt light-cone coordinates. Let $v = (t + x_1)/\sqrt{2}$, $w = (t - x_1)/\sqrt{2}$, and $x=x_2 $, $y=x_3 $, the null vector $k_\mu$ in $(v,w,x,y)$ coordinates is given by
\begin{equation}
k_\mu = 
\begin{pmatrix}
1,\frac{x^2+y^2}{2w^2+1},-\frac{2wx+\sqrt{2} y}{2w^2+1},-\frac{2wy-\sqrt{2} x}{2w^2+1}
\end{pmatrix}.
\end{equation}
Applying this null vector to the Kerr-Schild form, the vacuum Einstein equation determines the scalar function as \cite{Harada:2025krm}
\be
H=\frac{Nw}{2w^2+1}.
\ee
Restoring a dimensional parameter $b$ \cite{Harada:2025krm} and switching to the $(+,-,-,-)$ signature, the line-element of the solution is given by 
\be\label{metric}
\td s^2=2\td v \td w - \td x^2 - \td y^2 - H k_\mu k_\nu \td x^\mu \td x^\nu,
\ee
where 
\be
H=\frac{N w}{w^2+b^2},\qquad k_\mu \td x^\mu=\td v + \frac{x^2 + y^2 }{2(w^2+b^2)} \td w - \frac{by + wx}{w^2+b^2} \td x + \frac{bx - wy}{w^2+b^2} \td y.
\ee
We are now working in the $(+,-,-,-)$ signature which is commonly applied in the NP formalism. There are two free parameters $N$ and $b$. One can write the solution in the NP formalism where we choose the tetrad bases as
\be\label{Hopf}
\begin{split}
l&=\frac{x^2 + y^2 }{2(w^2+b^2)} \frac{\p}{\p v} + \frac{\p}{\p w} + \frac{wx + by }{w^2+b^2} \frac{\p}{\p x} + \frac{ wy - bx}{w^2+b^2} \frac{\p}{\p y},\\
n&=\frac{f(w,x,y)}{4(w^2+b^2)^2} \frac{\p}{\p v} + \frac{N w}{2(w^2+b^2)} \frac{\p}{\p w} + \frac{N w(wx + by)}{2(w^2+b^2)^2} \frac{\p}{\p x}  + \frac{N w (wy - bx) }{2(w^2+b^2)} \frac{\p}{\p y} ,\\
m&=-\frac{y + i x}{\sqrt{2} (w-ib)}\frac{\p}{\p v} - \frac{i}{\sqrt{2}} \frac{\p }{\p x} - \frac{1}{\sqrt{2}}\frac{\p }{\p y},
\end{split}
\ee
where 
\be
f(w,x,y)=4w^4+8b^2 w^2 + N w (x^2+y^2) + 4 b^4.
\ee
The basis vectors have the following orthogonality and normalization properties,
\be\label{tetradcondition}
l\cdot m=l\cdot\bm=n\cdot m=n\cdot\bm=0,\quad l\cdot n=1,\quad m\cdot\bm=-1.
\ee
The non-zero NP variables are obtained as
\be\label{HopfNP}
\rho=-\frac{1}{\chi},\qquad \chi=w-ib,\qquad\Psi_2=-\frac{N}{2\chi^3},\qquad 
\mu=\frac{N w }{2\xbar\chi\chi^2},\qquad \gamma=\frac{N}{4\chi^2}.
\ee
It is clear from the Weyl scalars that the solution is algebraic type-D. An exhaustive set of ten metrics for type-D was obtained in \cite{Kinnersley:1969zza}, see also comprehensive reviews in \cite{Stephani:2003tm} and references therein. Though the metric form in \eqref{metric} has not been presented elsewhere, it is reasonable to expect that the metric \eqref{metric} is diffeomorphic to a known solution.

From the NP variables in \eqref{HopfNP}, one can deduce that the imaginary part of $\Psi_2^0$, the leading piece of $\Psi_2$ in large $w$ expansion, vanishes and $b$ is the twist of the null congruence of $l$ in \eqref{Hopf}. Moreover, the $(x,y)$-surface at large $w$ has zero Gaussian curvature that corresponds to the $K=0$ case of the solutions in section 29.5 of \cite{Stephani:2003tm}. This can be verified from the large $w$ behavior of the spin coefficient $\mu$. Specifically, $\mu=\cO(w^{-2})$.

\section{Weyl invariants for type-D spacetime}

Before connecting the solution \eqref{metric} to other known solution, we first introduce some Weyl invariants for type-D solutions in this section. As a quick application, we use those Weyl invariants to distinguish the solution \eqref{metric} from the renowned type-D solution, the Kerr solution, which also has two free parameters.

For a algebraic type-D spacetime, one can always arrange the tetrad systems in NP formalism such that $l$ and $n$ are repeated principle null directions. Then the only non-zero Weyl scalar is $\Psi_2$ and the Goldberg-Sachs theorem \cite{Goldberg} yields that 
\be\label{Goldberg}
\kappa=\sigma=\lambda=\nu=0.
\ee
The Weyl tensor for type-D spacetime can be constructed from the Weyl scalars as\footnote{A factor $\frac12$ is missing for the first term on the right hand side in \cite{Chandrasekhar}.} 
\be\label{Weyl}
\begin{split}
W_{\mu\nu\alpha\beta}=&-\frac12(\Psi_2+\xbar\Psi_2)\Big[\{l_\mu n_\nu l_\alpha n_\beta\} + \{m_\mu \bm_\nu m_\alpha \bm_\beta\}\Big] \\
& + (\Psi_2 - \xbar\Psi_2) \{l_\mu n_\nu m_\alpha \bm_\beta\}  + \Psi_2 \{l_\mu m_\nu n_\alpha \bm_\beta\} + \xbar\Psi_2 \{l_\mu \bm_\nu n_\alpha m_\beta\} .
\end{split}
\ee
Note that the notation $\{\}$ means the permutation of the index as the Weyl tensor,
\be
\{\mu\nu\alpha\beta\}=(\mu \nu \alpha \beta - \nu \mu \alpha \beta - \mu \nu \beta \alpha + \nu \mu \beta \alpha + \alpha \beta \mu \nu - \beta \alpha \mu \nu -\alpha \beta \nu \mu  + \beta \alpha \nu \mu ).
\ee
The Weyl invariants of algebraic constructions can be obtained from the contractions of the tetrad system which depend only on the orthogonality and normalization conditions \eqref{tetradcondition}. A very efficient trick for the computation is that it is not necessary to use the tetrads from the solutions. One can use any four basis null vectors satisfying the same relations as \eqref{tetradcondition}. For instance, one can choose
\be
\label{simple}
l_\mu=(0,1,0,0),\quad n_\mu=(1,0,0,0),\qquad m_\mu=(0,0,0,1),\qquad \bm_\mu=(0,0,1,0).
\ee
The contractions are implemented by the fundamental matrix
\be
\eta^{\mu\nu}=\begin{pmatrix}
0 & 1 & 0 & 0\\
1 & 0 & 0 & 0\\
 0 & 0 & 0 & -1\\
 0 & 0 & -1 & 0
\end{pmatrix}.
\ee
Such choices will yield precisely the same results as choosing solution space tetrads and contracting by the spacetime metric when computing the Weyl invariants. Nonetheless, the computations are much more efficient, e.g., by using Mathematica. Some simpler Weyl invariants for type-D solutions are obtained as \cite{Lu:2025fzm}
\begin{align}
&W_{\mu\nu\alpha\beta}W^{\mu\nu\alpha\beta}=24\left[(\Psi_2)^2 + (\xbar\Psi_2)^2\right],\label{W2}\\
&{W^{\mu\nu}}_{\alpha\beta} {W^{\alpha\beta}}_{\rho\sigma} {W^{\rho\sigma}}_{\mu\nu}=48\left[(\Psi_2)^3 + (\xbar\Psi_2)^3\right],\label{W3}\\
&{W^{\mu\nu}}_{\alpha\beta} {W^{\alpha\beta}}_{\rho\sigma} {W^{\rho\sigma}}_{\tau\lambda}{W^{\tau\lambda}}_{\mu\nu}=288\left[(\Psi_2)^4 + (\xbar\Psi_2)^4\right].\label{W4}
\end{align}

The above algorithm that using simple non-solution tetrad bases to compute the Weyl invariants of algebraic construction can be extended to the calculation of invariants with covariant derivatives. In the NP formalism or other tetrad formalism, the tensor fields are projected onto the tetrad frame. The tensor fields are then obtained from the direct product of the tetrad components and the tetrad bases, such as the Weyl tensor in \eqref{Weyl}. Crucially, the tetrad components are scalar fields. The covariant derivative of the Weyl tensor,
\be
\begin{split}
\n_\tau W_{\mu\nu\alpha\beta}=&-\frac12(\n_\tau\Psi_2+\n_\tau\xbar\Psi_2)\Big[\{l_\mu n_\nu l_\alpha n_\beta\} + \{m_\mu \bm_\nu m_\alpha \bm_\beta\}\Big] \\
& -\frac12(\Psi_2+ \xbar\Psi_2)\n_\tau \Big[\{l_\mu n_\nu l_\alpha n_\beta\} + \{m_\mu \bm_\nu m_\alpha \bm_\beta\}\Big] \\
& + (\n_\tau\Psi_2 - \n_\tau\xbar\Psi_2) \{l_\mu n_\nu m_\alpha \bm_\beta\}  + (\Psi_2 - \xbar\Psi_2) \n_\tau \Big( \{l_\mu n_\nu m_\alpha \bm_\beta\}\Big)\\
&+ \n_\tau \Psi_2 \{l_\mu m_\nu n_\alpha \bm_\beta\} + \Psi_2 \n_\tau \Big( \{l_\mu m_\nu n_\alpha \bm_\beta\}\Big)\\
&+ \n_\tau \xbar\Psi_2 \{l_\mu \bm_\nu n_\alpha m_\beta\} +  \xbar\Psi_2 \n_\tau \Big(\{l_\mu \bm_\nu \n_\alpha m_\beta\}\Big),
\end{split}
\ee
involves two types of new terms. The first type is one covariant derivative acting on the Weyl scalar, which can be projected onto the tetrad frame by the directional derivatives,
\be\label{dP}
\n_\mu \Psi_2=D \Psi_2 n_\mu + \Delta \Psi_2 l_\mu - \delta \Psi_2 \bm_\mu - \bar\delta \Psi_2 m_\mu,
\ee
where $D,\Delta,\delta,\bar\delta$ are the  directional derivatives associated to the tetrads $l,n,m,\bm$, respectively. It is important to notice that any directional derivative of the Weyl scalar is also a scalar field. Then this part of the covariant derivative of Weyl tensor is again the direct product of scalar fields and the tetrad bases. The second type involves the covariant derivative acting on the tetrads, which can be projected onto the tetrad frame by the spin coefficients. The results with respect to the conditions \eqref{Goldberg} are
\begin{align}
\n_\nu l_\mu &= (\epsilon+\bar \epsilon)n_\nu l_\mu + (\gamma+\bar \gamma) l_\nu l_\mu -\tau l_\nu \bm_\mu - \bar \tau l_\nu m_\mu \nn \\
&\hspace{0.5cm} - (\beta+\bar \alpha) \bm_\nu l_\mu  + \bar \rho \bm_\nu m_\mu - (\bar \beta+ \alpha) m_\nu l_\mu  + \rho m_\nu \bm_\mu,\label{dL}\\
\n_\nu n_\mu &=-(\epsilon+\bar \epsilon) n_\nu n_\mu + \bar\pi n_\nu \bm_\mu + \pi n_\nu m_\mu   - (\gamma + \xbar\gamma) l_\nu n_\mu \nn \\
&\hspace{0.5cm}  + (\xbar\alpha + \beta) \bm_\nu n_\mu - \mu \bm_\nu m_\mu + (\alpha + \xbar\beta) m_\nu n_\mu - \bar\mu m_\nu \bm_\mu,\label{dN}\\
\n_\nu m_\mu &= (\epsilon - \bar\epsilon) n_\nu  m_\mu + \bar\pi n_\nu  l_\mu + (\gamma - \bar \gamma) l_\nu m_\mu - \tau l_\nu  n_\mu \nn \\
&\hspace{0.5cm} - (\beta - \xbar\alpha) \bm_\nu m_\mu  - (\alpha - \xbar \beta) m_\nu m_\mu + \rho m_\nu n_\mu - \xbar\mu m_\nu l_\mu. \label{dM} 
\end{align}
Now the covariant derivative of Weyl tensor $\n_\tau W_{\mu\nu\alpha\beta}$ has been arranged in a way that it is a product of scalar fields and tetrad bases. Again, the contractions of the tensorial index involve only the contractions of different tetrad bases. Then, one can apply the same strategy for computing Weyl invariants involving one covariant derivative, namely choosing simple non-solution bases. For type-D spacetime, one obtains the following type of Weyl invariant with one covariant derivative \cite{Lu:2025fzm}
\be
\n_\tau W_{\mu\nu\alpha\beta} \n^\tau W^{\mu\nu\alpha\beta}= 240 \left(\rho \Psi_2 \Delta \Psi_2 + \xbar\rho \xbar\Psi_2 \Delta \xbar\Psi_2- \tau \Psi_2 \xbar\delta \Psi_2  - \xbar\tau \xbar\Psi_2 \delta \xbar\Psi_2\right). 
\ee
One can apply the Bianchi identity
\be
D \Psi_2 = 3\rho \Psi_2,\quad \Delta \Psi_2 =-3\mu \Psi_2,\quad \delta\Psi_2=3\tau \Psi_2,\quad \xbar\delta \Psi_2 = - 3\pi \Psi_2,
\ee
to arrange this invariant as
\be\label{dW2}
\n_\tau W_{\mu\nu\alpha\beta} \n^\tau W^{\mu\nu\alpha\beta}=-720\bigg[ (\Psi_2)^2(\mu \rho -  \pi \tau) + (\xbar\Psi_2)^2 ( \bar\mu \bar\rho -  \bar\pi \bar\tau)
\bigg].
\ee

Now We will apply the Weyl invariants $W_{\mu\nu\alpha\beta}W^{\mu\nu\alpha\beta}$, ${W^{\mu\nu}}_{\alpha\beta} {W^{\alpha\beta}}_{\rho\sigma} {W^{\rho\sigma}}_{\mu\nu}$, and $\n_\tau W_{\mu\nu\alpha\beta} \n^\tau W^{\mu\nu\alpha\beta}$ to show that the solution in \eqref{metric} and the Kerr solution are distinct solutions of vacuum Einstein equation.\footnote{For type-D solutions, ${W^{\mu\nu}}_{\alpha\beta} {W^{\alpha\beta}}_{\rho\sigma} {W^{\rho\sigma}}_{\tau\lambda}{W^{\tau\lambda}}_{\mu\nu}$ is not an independent variant to distinguish solutions as it can be expressed in terms of $W^2$ and $W^3$.} We denote those three invariants as $W^2$, $W^3$, and $dW^2$, respectively. For the solution \eqref{metric}, those three Weyl invariants are obtained as
\be
\begin{split}
&W^2=6 N^2 \left(\frac{1}{\chi^6} + \frac{1}{\xbar\chi^6}\right),\qquad W^3=6 N^3 \left(\frac{1}{\chi^9} + \frac{1}{\xbar\chi^9}\right),\\ 
&dW^2=45 N^3 \left(\frac1\chi + \frac{1}{\xbar\chi}\right) \left(\frac{1}{ \chi^8} + \frac{1}{ \xbar\chi^8}\right).
\end{split}
\ee
It is obvious that the Weyl invariants are determined by three variables $N$, $\chi$, and $\xbar\chi$. However, the parameter $N$ has a different nature from $\chi,\xbar\chi$. Actually, $N$ can be absorbed by the rescaling of variables $\tilde\chi=\frac{\chi}{N^{\frac13}}$ and $\xbar{\tilde\chi}=\frac{\xbar\chi}{N^{\frac13}}$. The three Weyl invariants with the rescaled variables are given by
\be\label{Hopf-Weyl}
\begin{split}
&W^2=6 \left(\frac{1}{\tilde\chi^6} + \frac{1}{\xbar{\tilde\chi}^6}\right),\qquad W^3=6  \left(\frac{1}{\tilde\chi^9} + \frac{1}{\xbar{\tilde\chi}^9}\right),\\ 
&dW^2=45 \left(\frac{1}{\tilde\chi} + \frac{1}{\xbar{\tilde\chi}}\right) \left(\frac{1}{ \tilde\chi^8} + \frac{1}{ \xbar{\tilde\chi}^8}\right).
\end{split}
\ee
Now there are two variables left for those three Weyl invariants. In principle, one can eliminate $\tilde\chi$ and $\xbar{\tilde\chi}$ by $W^2$ and $W^3$. Then a unique relation among $W^2$, $W^3$, and $dW^2$ can be recovered for the solution \eqref{metric}.

If two metrics are the same solution related by coordinates transformation, the relation of the Weyl invariants must be the same. Conversely, if one can prove that the unique relation of Weyl invariants for one solution is not satisfied for the other solution, those two solution must be distinct solutions. In the next subsection, We will demonstrate that the metric in \eqref{metric} and the Kerr metric are distinct solutions using the relation revealed from \eqref{Hopf-Weyl}.

\subsection{Kerr solution}

In Boyer-Linquist coordinates $(t,r,\theta,\phi)$, the tetrads of Kerr solution in the NP formalism are given by \cite{Chandrasekhar}
\begin{equation}
\begin{split}
  &l = \frac{r^2+a^2}{\Delta} \frac{\p}{\p t} + \frac{\p}{\p r} + \frac{a}{\Delta} \frac{\p}{\p \phi},\qquad
  n = \dfrac{r^2+a^2}{2\varrho\bar\varrho}\frac{\p}{\p t} - \frac{\Delta}{2\varrho\bar\varrho}\frac{\p}{\p r} + \frac{a}{2\varrho\bar\varrho}\frac{\p}{\p \phi},\\
  &  m = \dfrac{i a \sin\theta}{\sqrt 2 \varrho}\frac{\p}{\p t} + \dfrac{1}{\sqrt 2 \varrho}\frac{\p}{\p \theta} + \frac{i}{\sqrt 2\sin\theta \varrho}\frac{\p}{\p \phi},
\end{split}
\end{equation}
and the NP variables are given by
\begin{equation}\label{Kerr}
\begin{split}
    & \rho=-\frac{1}{\bar\varrho},\quad
    \beta=\frac{\cot\theta}{\varrho\, 2\sqrt 2}, \quad\tau
    =-\frac{ia\sin\theta}{\varrho\bar\varrho\sqrt 2},\quad
    \pi=\frac{ia\sin\theta}{(\bar\varrho)^2\sqrt 2},\\
    & \mu=-\frac{\vartriangle}{2\varrho\bar\varrho^2},\quad
    \gamma=\mu+\frac{r-M}{2\varrho\bar\varrho},\quad \alpha=\pi-\bar\beta, \quad \Psi_2=-\frac{M}{(\bar\varrho)^3},
\end{split}
\end{equation}
where
\begin{equation}
  \vartriangle=r^2-2Mr+a^2, \quad
 \varrho= r+ia\cos\theta,\quad \bar\varrho=r-ia\cos\theta.
\end{equation}
Inserting the NP variables into the Weyl invariants introduced previously, we obtain
\be
\begin{split}
&W^2=24 M^2 \left[\frac{1}{\varrho^6} + \frac{1}{\xbar\varrho^6}\right],\qquad W^3=48 M^3 \left[\frac{1}{\varrho^9} + \frac{1}{\xbar\varrho^9}\right],\\ 
&dW^2=360 M^3 \left[\frac{1}{\varrho^8} + \frac{1}{\xbar\varrho^8}\right] \left[\left(\frac{1}{\varrho} + \frac{1}{\xbar\varrho}\right) - \frac{1}{M} \right].
\end{split}
\ee
It is clear that the Weyl invariants are determined by three variables $M$, $\varrho$, and $\xbar\varrho$. One would consider a similar rescaling of variables as the case for the solution \eqref{metric}. For instance, one can define $\tilde\varrho=\frac{\varrho}{(2M)^\frac13}$ and $\xbar{\tilde\varrho}=\frac{\xbar\varrho}{(2M)^\frac13}$. Then the Weyl invariants are given by
\be
\begin{split}
&W^2=6 \left[\frac{1}{\tilde\varrho^6} + \frac{1}{\xbar{\tilde\varrho}^6}\right],\qquad W^3=6 \left[\frac{1}{\tilde\varrho^9} + \frac{1}{\xbar{\tilde\varrho}^9}\right],\\ 
&dW^2=45 \left[\frac{1}{\tilde\varrho^8} + \frac{1}{\xbar{\tilde\varrho}^8}\right] \left(\frac{1}{\tilde\varrho} + \frac{1}{\xbar{\tilde\varrho}}\right) -  \frac{90}{(4M^2)^\frac13} \left[\frac{1}{\tilde\varrho^8} + \frac{1}{\xbar{\tilde\varrho}^8}\right].
\end{split}
\ee
Clearly, there are extra terms in $dW^2$ compared to the one in \eqref{Hopf-Weyl} for the solution \eqref{metric}. The extra terms involve ``new'' variable $M$ in the sense that $dW^2$ is not a function of variables ${\tilde\varrho}$ and $\xbar{\tilde\varrho}$ only. Hence, there can not a unique relation among $W^2$, $W^3$, and $dW^2$ for the Kerr solution, which is in contrast with the solution \eqref{metric}. The discrepancy for the relation among the Weyl invariants $W^2$, $W^3$, and $dW^2$ undoubtedly proves that the solution in \eqref{metric} and the Kerr solution are distinct solutions.

Before closing this section, we will comment on another difference between the solution in \eqref{metric} and the Kerr solution. As pointed out in previous section, the asymptotic boundary of \eqref{metric} at large $w$ has zero Gaussian curvature, namely the boundary is a plane. The asymptotic boundary of the Kerr solution is a sphere. However, the difference of the asymptotic boundary does not necessarily imply that those two solutions are distinct. A counter example is that the Minkowski spacetime can be written in both forms \cite{Mao:2024pqz}. More precisely, the Minkowski spacetime with a sphere boundary in retarded stereographic coordinates $(u_s,r_s,z_s ,\bz_s)$ is given by
\be\label{sphere}
d s^2=d u_s^2+ 2d u_s d r_s 
 - \frac{2r_s^2}{P_s^2} d z_s d\bz_s,\qquad P_s=\frac{1+z\bz}{\sqrt2}.
\ee
The following change of coordinates 
\be
\begin{split}
&r_s=\frac{1}{\sqrt2} \sqrt{[u_p + r_p(1+z_p \bz_p)]^2-4 r_p u_p},\\
&u_s=\frac{1}{\sqrt2} [u_p + r_p(1+z_p \bz_p)] - r_s,\\
&z_s=\frac{\frac{1}{\sqrt2} [u_p - r_p(1+z_p \bz_p)]+r_s}{u_s \bz_p},
\end{split}
\ee
can transform the spacetime to the case with a plane boundary, where the line-element is given by
\be\label{plane}
d s^2=2d u_p d r_p - 2r_p^2 d z_p d\bz_p.
\ee

\section{Type-D solution with plane boundary}

As pointed out previously, the solution \eqref{metric} is expected to be diffeomorphic to a known type-D solution. In this section, we search from the complete set of known type-D solutions, e.g., in \cite{Kinnersley:1969zza}, for this solution. One can directly select the type-D solution that fulfills the same properties as \eqref{HopfNP} from the generic solution space with a plane asymptotic boundary in \cite{Kinnersley:1969zza,Stephani:2003tm}, see also \cite{Lu:2025fzm} for more relevant information. In $(u,r,z,\bz)$ coordinates, the solution in the NP formalism corresponding to \eqref{Hopf} is given by 
\be\label{Our}
l=\frac{\p}{\p r},\qquad n=\frac{\p}{\p u} + \frac{N r}{2(r^2 + b^2)}\frac{\p}{\p r},\qquad m=-\frac{ib z}{r+ib} \frac{\p}{\p u} - \frac{1}{r+ib}\frac{\p}{\p \bz},
\ee
which yields the line-element as
\be
\begin{split}\label{Ourmetric}
\td s^2=&2\td u \td r - 2 r^2 \td z \td \bz \\
&-\frac{r N}{r^2+b^2}\td u^2 - \frac{2 i N b r \bz }{r^2+b^2} \td u \td z + \frac{2 i N b r z }{r^2+b^2} \td u \td \bz + 2 i b \bz \td r \td z \\
&- 2 i b z \td r \td \bz + \frac{N b^2 r \bz^2}{r^2+b^2}\td z^2 + \frac{N b^2 r z^2}{r^2+b^2}\td \bz^2 - 2b^2\left(1 + \frac{N r z \bz}{r^2+b^2} \right) \td z \td \bz.
\end{split}
\ee
The non-zero NP variables are given by 
\be\label{OurNP}
\rho=-\frac{1}{\varrho},\qquad \quad \varrho=r-ib, \qquad \Psi_2=-\frac{N}{2\varrho^3},\qquad \mu=\frac{r N}{2\xbar\varrho\varrho^2} \quad \quad \gamma=\frac{N}{4\varrho^2}.
\ee
The above metric is given in completely distinct forms compared to the one in \eqref{metric}. But the NP variables are very similar. Actually every individual NP variable from the two solutions is given in exactly the same function of different variables. They are identical if one identifies the variables $w=r$. Hence, all Weyl invariants of algebraic construction from the NP variables for the two metrics must have exactly the same form. As an example, one can check the Weyl invariants studied previously, namely, 
\be
\begin{split}
&W^2=6 N^2 \left(\frac{1}{\varrho^6} + \frac{1}{\xbar\varrho^6}\right),\qquad W^3=6 N^3 \left(\frac{1}{\varrho^9} + \frac{1}{\xbar\varrho^9}\right),\\ 
&dW^2=45 N^3 \left(\frac1\varrho + \frac{1}{\xbar\varrho}\right) \left(\frac{1}{ \varrho^8} + \frac{1}{ \xbar\varrho^8}\right).
\end{split}
\ee
Clearly, the relation among $W^2$, $W^3$, and $dW^2$ for the solution \eqref{Ourmetric} must be identical to the solution \eqref{metric}. This is a very positive indication that the two solutions are the same. Nevertheless, this is not enough. One needs to also check Weyl invariants with all possible covariant derivatives acting on the NP variables.

The new ingredients for computing those types of Weyl invariants are the spin coefficients acted by one or more covariant derivatives and the Weyl scalar acted by two or more covariant derivatives. They must contribute identically to the Weyl invariants following precisely the justification for the directional derivatives of the Weyl scalar $\Psi_2$. More precisely, let us make the identification $w=r$, which yields that the NP variables in \eqref{HopfNP} and \eqref{OurNP} are identical. A covariant derivative acting on any NP variable can be projected on the null frame as
\be
\n_\mu S=D S n_\mu + \Delta S l_\mu - \delta S \bm_\mu - \xbar\delta S m_\mu,
\ee
where $S$ denotes any NP variable. Since the tetrads \eqref{Hopf} and \eqref{Our} are in completely different forms, the directional derivative would give rise to distinctions for the two solutions. Interestingly, the NP variables in \eqref{HopfNP} and \eqref{OurNP} are only functions of $w$ and $r$, respectively. That means only the $w$-component of \eqref{Hopf} and $r$-component of \eqref{Our} are relevant for computing the covariant derivative, which are identical with the identification $w=r$. Hence, all NP variables acted by one covariant derivative are identical for the two solutions up to the tetrad bases. Since one needs to compute the contractions for all tetrad bases for deriving the Weyl invariants, one can conclude that the contributions from all NP variables acted by one covariant derivative are identical. Moreover, $w$-component of \eqref{Hopf} and $r$-component of are only functions of $w$ and $r$, respectively. The above algorithm can be extended to the case with arbitrary covariant derivatives acting on the NP variables. That means all possible Weyl invariants for the two solution must be identical. Thus, any relations of the Weyl invariants must be the same for those two metrics. As both solutions are vacuum solution of four-dimensional Einstein equation, the identification of the Weyl invariants yields that they must be the same solution, at least locally.

\subsection{Coordinates transformation}

To close this section, we will present the precise coordinates transformation that connecting the two metrics \eqref{metric} and \eqref{Ourmetric}. In general, finding the direct connection of the two coordinates systems for two completely distinct metrics is not easy. Fortunately, the identification $w=r$ will simplify the derivation. Moreover, the tetrad bases $l$ and $n$ for both cases \eqref{Hopf} and \eqref{Our} are repeated principal null directions. So one can derive the coordinates transformation from the null bases which yields first order partial differential equations and significantly simplifies the derivation. The precise coordinates transformation is
\be
u=v-\frac{w(x^2+y^2)}{2(w^2+b^2)},\qquad r=w,\qquad z=\frac{wy-bx}{\sqrt{2}(w^2+b^2)}+i\frac{wx+by}{\sqrt{2}(w^2+b^2)},
\ee
which undoubtedly verifies that the two metrics are diffeomorphic to each other.

\section{Conclusions}

In this note, we use Weyl invariants to demonstrate that two distinct metrics actually represent the same vacuum solution of Einstein equation. While, this serves as a single concrete example, it naturally raises the question of how generally applicable the present algorithm is. The key point of this work lies in the identification of exact solutions through Weyl invariants. In fact, this approach can be regarded as the last step in the study of exact solutions. Once a solution is obtained, one typically studies its geometric features, such as its algebraic classification or singularity structure. These properties already provide strong means of distinguishing different solutions.

When two solutions fall within the same subclass, for instance, type-D and no curvature singularity, the Weyl invariants constructed in this work offer a particularly effective tool for distinguishing or identifying them. Of course, it is not always practical to exhaust all possible Weyl invariants in order to identify solutions. Nevertheless, identifying partial relations among a subset of Weyl invariants can still provide valuable information to determine part of the coordinate correspondences between the two metrics, such as the identification $w=r$ in the present work.

\section*{Acknowledgments}

The author thanks Glenn Barnich, Hong L\"u, and Weicheng Zhao for useful discussions and collaborations in relevant research topics.
This work is supported in part by the National Natural Science Foundation of China (NSFC) under Grants No.~12475059 and No.~11935009, and by Tianjin University Self-Innovation Fund Extreme Basic Research Project Grant No. 2025XJ21-0007.


\providecommand{\href}[2]{#2}\begingroup\raggedright\endgroup

\end{document}